\documentclass[
superscriptaddress,
reprint,floatfix,
amsmath,amssymb,
aps,
prapplied,
longbibliography
]{revtex4-2}
\usepackage{color}
\usepackage{graphicx}
\usepackage{bm}
\usepackage{mathrsfs}
\usepackage{dsfont}
\usepackage{tikz}
\usetikzlibrary{decorations.pathmorphing}
\usepackage[unicode=true, breaklinks=false, pdfborder={0 0 1}, backref=false, colorlinks=true, linkcolor=blue, citecolor=blue]{hyperref}
\usepackage{physics}

\graphicspath{{./}}

\begin{document}
\title{Double-quantum enhanced coherent control and magnetic sensing of spin defects in hexagonal boron nitride}

\author{Shihao Ru}
\affiliation{School of Electrical and Electronic Engineering, Nanyang Technological University, Singapore}
\affiliation{Division of Physics and Applied Physics, School of Physical and Mathematical Sciences, Nanyang Technological University, Singapore 637371, Singapore}
\affiliation{Centre for Quantum Technologies, Singapore}

\author{Abdullah Rasmita}
\affiliation{School of Electrical and Electronic Engineering, Nanyang Technological University, Singapore}
\affiliation{Division of Physics and Applied Physics, School of Physical and Mathematical Sciences, Nanyang Technological University, Singapore 637371, Singapore}

\author{Jiakai Chen}
\affiliation{Centre for Quantum Technologies, Singapore}

\author{Yuzhe Yang}
\affiliation{Division of Physics and Applied Physics, School of Physical and Mathematical Sciences, Nanyang Technological University, Singapore 637371, Singapore}

\author{Xiaodan Lyu}
\affiliation{School of Electrical and Electronic Engineering, Nanyang Technological University, Singapore}

\author{Robert Cernansky}
\affiliation{Institute for Quantum Optics, Ulm University, Albert-Einstein-Allee 11, 89081 Ulm, Germany}

\author{Edwin Hang Tong Teo}
\affiliation{School of Electrical and Electronic Engineering, Nanyang Technological University, Singapore}

\author{Jes\'us Z\'u\~niga-Pérez}
\affiliation{Division of Physics and Applied Physics, School of Physical and Mathematical Sciences, Nanyang Technological University, Singapore 637371, Singapore}
\affiliation{Majulab International Joint Research Unit UMI 3654, CNRS, Université Côte d’Azûr, Sorbonne Université, National
University of Singapore, Nanyang Technological University, Singapore, Singapore}

\author{Fedor Jelezko}
\affiliation{Institute for Quantum Optics, Ulm University, Albert-Einstein-Allee 11, 89081 Ulm, Germany}

\author{Weibo Gao}
\email[Contact author: ]{wbgao@ntu.edu.sg}
\affiliation{School of Electrical and Electronic Engineering, Nanyang Technological University, Singapore}
\affiliation{Division of Physics and Applied Physics, School of Physical and Mathematical Sciences, Nanyang Technological University, Singapore 637371, Singapore}
\affiliation{Centre for Quantum Technologies, Singapore}
\affiliation{Quantum Science and Engineering Centre (QSec), Nanyang Technological University, Singapore, Singapore}

\begin{abstract}
The negatively charged boron vacancy ($V_{\mathrm{B}}^{-}$) in hexagonal boron nitride (hBN) is a promising spin defect for quantum sensing in two-dimensional material systems.
Previous hBN experiments have mainly exploited the two spin-triplet ground-state single-quantum transitions separately. Here we implement a double-quantum driving scheme based on simultaneous resonant microwave excitation of both ground-state single-quantum transitions, establishing coherent control of the $\Delta m_s=2$ manifold in a $V_{\mathrm{B}}^{-}$ ensemble.
We compare the resulting double-quantum response with the corresponding single-quantum protocol under identical experimental conditions. The double-quantum protocol enhances the optical readout contrast and yields a 2.6-fold signal-to-noise improvement in AC magnetometry.
We further use the double-quantum relaxation channel as a differential probe of electric-field and strain noise, showing that the present data are compatible with correlated-surface or volume-distributed fluctuations, while short-range uncorrelated surface-noise models are disfavored within the assumptions of our analysis.
These results establish double-quantum coherent control and frequency-resolved AC magnetometry in hBN spin ensembles, and identify the sample-dependent measurements needed to determine the microscopic origin of the dominant relaxation channels.
\end{abstract}

\maketitle

\section{Introduction}

Two-dimensional van der Waals materials offer a new platform for hosting solid-state spin defects, enabling scalable quantum devices beyond conventional bulk systems. Among them, hexagonal boron nitride (hBN) stands out due to its wide bandgap, robust thermal and mechanical properties, and optically addressable spin defects
\cite{caldwell2019photonics,liu2022spin,gao2025single,robertson2025charge}.
Specifically, the negatively charged boron vacancy ($V_{\mathrm{B}}^{-}$) introduces deep-level electronic states in the wide-bandgap hBN host that exhibit spin-dependent intersystem crossing, enabling optical initialization and readout of its spin state at room temperature 
\cite{gottscholl2020initialization,gottscholl2021room,Vaidya31122023,fang2024quantum,dlzw-dhsr}. Its electronic structure is characterized by a spin-triplet ground state ($\lvert m_s = 0, \pm 1 \rangle$), which provides a robust basis for coherent spin manipulation under ambient conditions \cite{Mu_prl_2022,jiangfeng_nano_esodmr,mathur2022excited,stern2024quantum,Ruprl2024}.
As a result, the $V_{\mathrm{B}}^{-}$ exhibits strong coupling to magnetic fields \cite{zhou_nanoletter}, strain and electric fields \cite{Lyu_nanoletter_strain,zhou2025detection,mu2025spin}, and temperature \cite{gottscholl2021spin}, making it well suited for high-sensitivity quantum sensing at the nanoscale.
This capability is further enhanced in isotopically optimized hBN, where the reduced nuclear-spin bath suppresses magnetic noise and enables extended spin coherence time \cite{Sasaki_2023,PhysRevLett.131.126901,gong2024isotope,PhysRevLett.134.096202}.

To harness these properties for practical sensing applications, color-center–based magnetometry techniques have been developed to coherently manipulate and read out the spin states of such defects \cite{RevModPhys.89.035002,RevModPhys.92.015004,RevModPhys.96.025001}. Conventional approaches, including continuous-wave optically detected magnetic resonance (cw-ODMR), pulsed ODMR, and dynamical-decoupling protocols \cite{PhysRevLett.133.166704,rizzato2023extending}, are typically implemented within the pseudo-spin-$1/2$ single-quantum (SQ) manifold of the ground state. For the $V_{\mathrm{B}}^{-}$ defect, this manifold is defined by the $\lvert m_s = 0 \rangle$ state and one of the $\lvert m_s = \pm 1 \rangle$ states, with the $\Delta m_s = 1$ transition serving as the sensing channel \cite{doi:10.1126/sciadv.adk8495,doi:10.1021/acsphotonics.5c00891,huang2022wide}.
In contrast, double-quantum (DQ) coherence magnetometry ($\Delta m_s = 2$) exploits the full spin-1 nature of the $V_{\mathrm{B}}^{-}$ defect and offers several intrinsic advantages for magnetic sensing. In particular, by simultaneously driving both $\lvert 0 \rangle \leftrightarrow \lvert -1 \rangle$ ($f_-$) and $\lvert 0 \rangle \leftrightarrow \lvert +1 \rangle$ ($f_+$) transitions, the magnetic phase accumulation is doubled relative to an SQ coherence.
This concept of concurrent excitation of multiple electron-spin transitions has previously been demonstrated in nitrogen-vacancy (NV) centers in diamond \cite{PhysRevLett.113.027602,PhysRevLett.118.197201,PhysRevApplied.15.054032, doi:10.1073/pnas.1306825110,PhysRevApplied.15.044020,lourette2026temperatureinsensitivecompositediamondclock,jarmola2021demonstration} and silicon-vacancy centers in silicon carbide \cite{PhysRevApplied.19.044086,tahara2025quantum}. The contribution of the present work is to establish this differential probe in $V_{\mathrm{B}}^{-}$ ensembles in hBN, where the layered host geometry, dense nuclear-spin bath, and strong ensemble broadening create a distinct experimental regime.
\begin{figure*}[!htbp]
  \centering
  \includegraphics[width=0.9\textwidth]{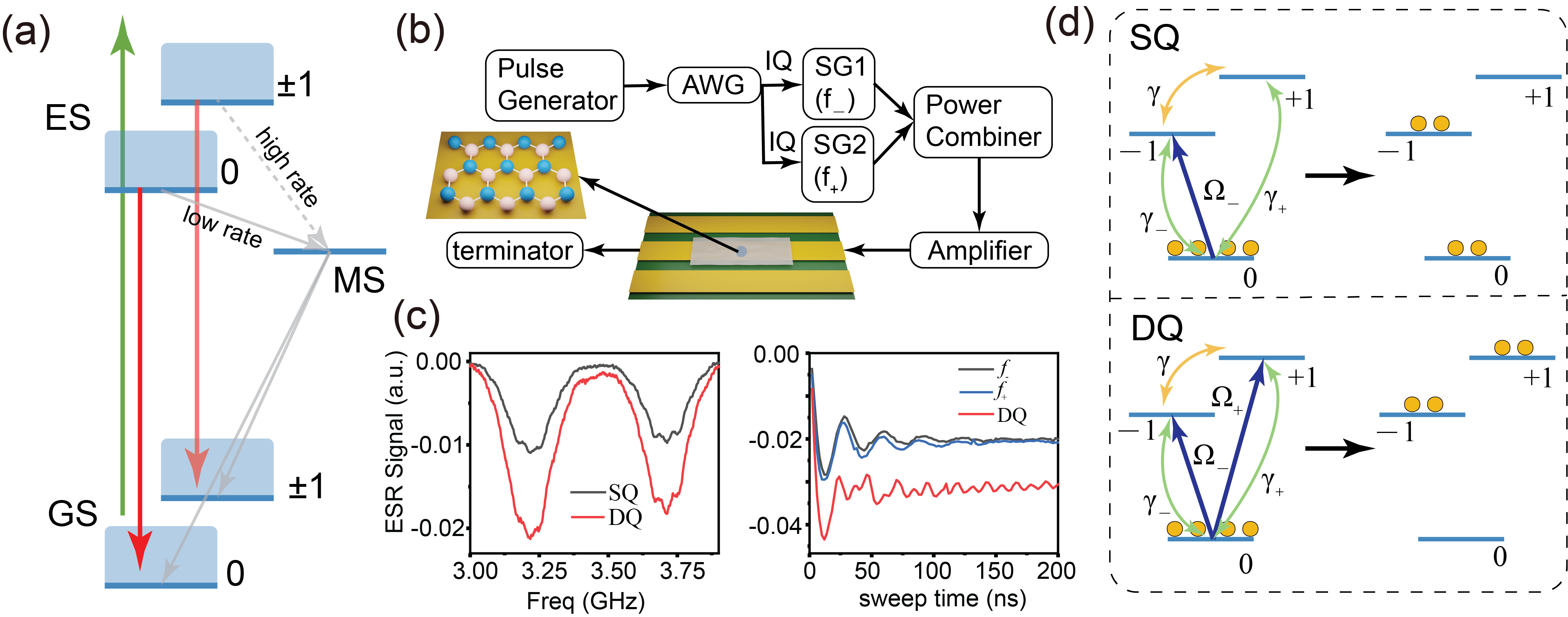}
  \caption{Double-quantum driving scheme for the $V_{\mathrm{B}}^{-}$ defect.
  (a) Energy-level structure of $V_{\mathrm{B}}^{-}$ in hBN.
  (b) Microwave signal generation scheme for double-quantum driving.
  (c) ODMR and coherent Rabi oscillations measured under double-quantum driving. The black and blue traces correspond to single-quantum (SQ) measurements on the $\lvert 0 \rangle \leftrightarrow \lvert \pm 1 \rangle$ transitions, while the red trace represents the double-quantum (DQ) response obtained by simultaneously driving both transitions.
  (d) Triplet-qubit operation scheme: a non-resonant laser pulse initializes the spin into the $\lvert 0 \rangle$ state, microwave ($\pi$) pulses drive transitions between $\lvert 0 \rangle$ and $\lvert \pm 1 \rangle$, and spin-dependent fluorescence enables optical readout.
  }
\label{fig:dq-driving}
\end{figure*}

In this work, we demonstrate room-temperature control of the full spin-triplet structure of negatively charged boron vacancies in hexagonal boron nitride. We introduce a microwave pulse scheme that enables the simultaneous and coherent driving of the two single-quantum transitions, thereby realizing control of the $\Delta m_s = 2$ spin coherence. The resulting spin dynamics are investigated using ODMR and Rabi measurements, and the corresponding spin relaxation rates are extracted under various magnetic fields. A quantitative analysis of the DQ relaxation channel constrains, but does not uniquely identify, the spatial origin of the field-dependent electric-field/strain noise. Building on this control capability, we apply this approach to AC magnetometry using spin-echo correlation spectroscopy and quantum heterodyne readout, demonstrating room-temperature detection of external AC magnetic fields.
\section{Experimental implementation of DQ driving}
At room temperature, the ground-state spin of the $V_{\mathrm{B}}^{-}$ defect is initialized using a green laser pulse of approximately $9~\mu\mathrm{s}$, as illustrated in Fig.~\ref{fig:dq-driving}(a). While the radiative decay rates of the $\lvert m_s = 0 \rangle$ and $\lvert m_s = \pm 1 \rangle$ excited states are comparable and rather small, their nonradiative decay rates through the metastable state are orders of magnitude larger but differ crucially by a factor $\sim$2. The full relaxation dynamics results in spin-dependent fluorescence with the contribution of the $\lvert m_s = \pm 1 \rangle$ states to the overall fluorescence being "darker" than the contribution of the $\lvert m_s = 0 \rangle$ state, giving rise to a negative ODMR contrast \cite{PhysRevB.110.014104}.
Before presenting the results, we briefly outline the microwave control scheme employed in this work, as shown in Fig. \ref{fig:dq-driving}(b).
All instruments are synchronized to a common 10 MHz clock, which serves as the timebase reference.
The microwave signal generators are operated exclusively in the IQ (In-phase and Quadrature) modulation mode (see Appendix~\ref{app:microwave}), with sideband frequency and phase control implemented using IQ waveforms generated by an arbitrary waveform generator (AWG).
After IQ modulation, the residual local-oscillator carrier is suppressed by approximately 50 dB relative to the modulated output.
To prevent the residual LO signals from the I/Q mixers from unintentionally driving the system, we strategically assigned the sideband polarities: Signal Generator 1 is set to the blue sideband to address $f_-$, and Signal Generator 2 to the red sideband for $f_+$. This configuration ensures the LO frequencies remain outside the relevant transition spectrum, given $f_- < f_+$; their outputs are combined using a microwave power combiner, preserving phase coherence.

The hBN flake is transferred onto a coplanar waveguide (CPW) patterned on a 200-nm-thick Au film. The microwave signals are delivered to the CPW to generate a strong local AC magnetic field at the position of the hBN flake, which drives the $\lvert 0 \rangle \leftrightarrow \lvert \pm 1 \rangle$ transitions.
For all ODMR measurements, laser on/off modulation, as well as the transistor--transistor logic (TTL) trigger signals required by the detection hardware and the arbitrary waveform generator (AWG), are generated by a pulse generator.
In addition, a three-dimensional Helmholtz coil is used to generate the magnetic field, providing each uniaxial static DC magnetic field tunable between $\pm12$ mT.

The timing sequences used for pulsed ODMR and Rabi oscillation measurements are similar.
In pulsed ODMR experiments, the microwave pulse duration is fixed at $400$~ns while the microwave frequency is swept, allowing the resonance frequencies ($f_{\pm}$) to be identified from the resulting spectral response.
These resonance frequencies are then used for Rabi oscillation measurements, in which the microwave frequency is fixed at $f_{\pm}$ and the microwave pulse duration is swept.
To improve the signal-to-noise ratio (SNR), the signal is integrated using a lock-in amplifier with a $100$~ms time constant.
Representative pulsed ODMR and Rabi oscillation spectra acquired under these conditions are shown in Fig.~\ref{fig:dq-driving}(c).
At a fixed magnetic field, spins prepared in a coherent superposition of the $\lvert +1 \rangle$ and $\lvert -1 \rangle$ states precess at twice the rate of conventional SQ coherences, $\{\lvert 0 \rangle, \lvert \pm 1 \rangle\}$, as illustrated schematically in Fig.~\ref{fig:dq-driving}(d).
DQ-based measurements are also inherently differential: perturbations common to the two $\lvert 0 \rangle \leftrightarrow \lvert \pm 1 \rangle$ transitions are suppressed, whereas transverse electric-field and strain fluctuations that couple the $\lvert +1 \rangle$ and $\lvert -1 \rangle$ states are retained in the DQ relaxation channel.
Experimentally, the most direct advantage in the present ensemble is the optical readout contrast, which is approximately twice as large as in the SQ scheme. For an isolated spin the DQ drive can also yield a $\sqrt{2}$ enhancement of the Rabi frequency, but in our ensemble the inhomogeneous detuning makes the dynamics more complex and the robust advantage is the measured contrast and SNR improvement (see Appendix~\ref{app:simulation}).
\begin{figure}[!htbp]
  \centering
  \includegraphics[width=\linewidth]{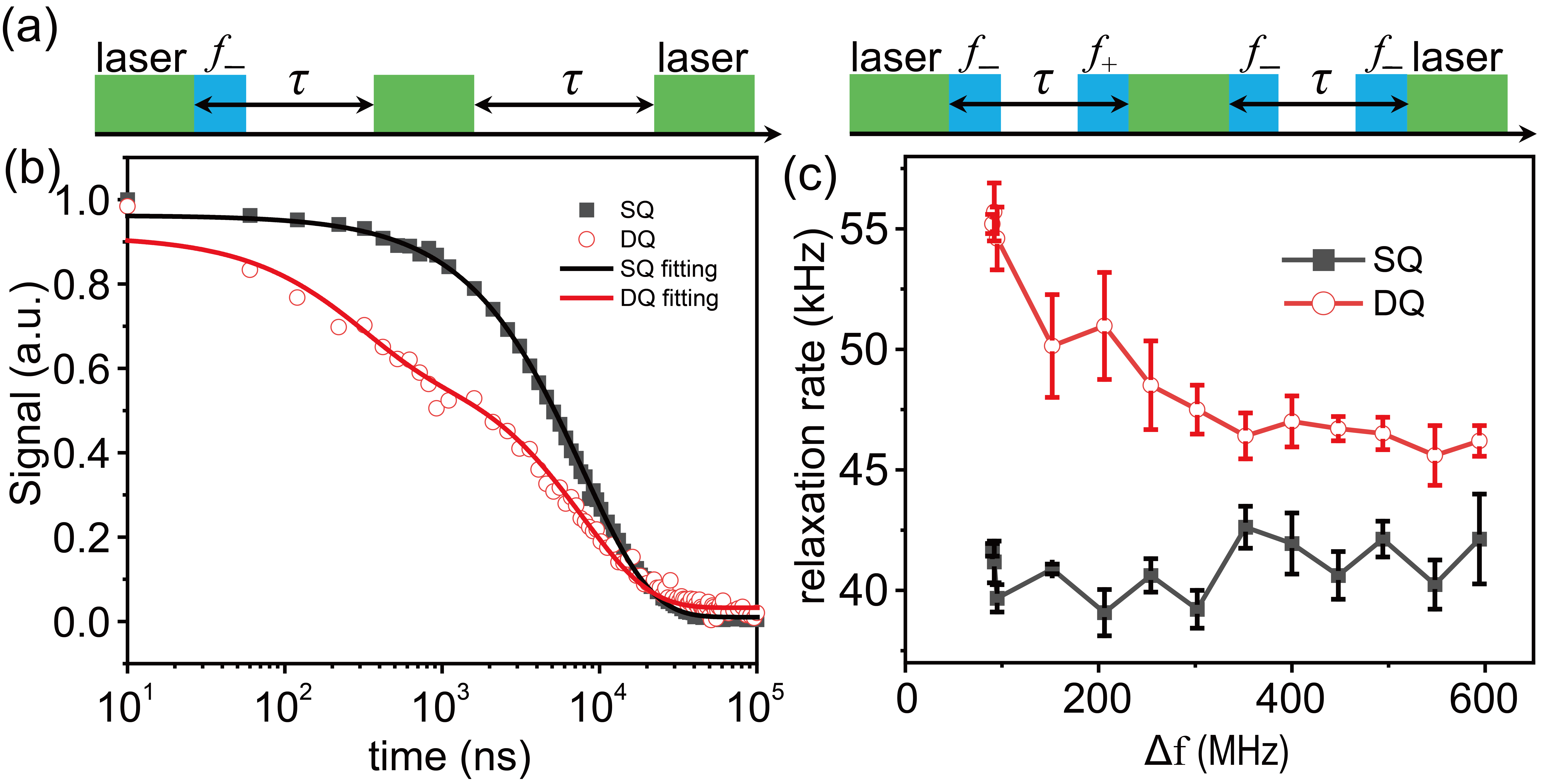}
  \caption{Double- and single-quantum relaxation rates.
  (a) Pulse sequences used to measure the DQ relaxation rate ($\gamma$) and the SQ relaxation rate ($\gamma_0$). Green laser pulses initialize and read out the $V_{\mathrm{B}}^{-}$ ground-state spin. The first sequence corresponds to a $T_1$ measurement with a single microwave $\pi_{0,-1}$ pulse. For DQ relaxation measurements, the system evolves during a dark interval $\tau$, after which the population of the $\lvert \pm 1 \rangle$ states is mapped onto the $\lvert 0 \rangle$ state using a microwave $\pi_{0,\pm 1}$ pulse prior to detection.
  (b) Population signal with fits to a three-level relaxation model (solid lines).
  (c) Magnetic-field dependence of the DQ relaxation rate $\gamma$, plotted against the ODMR splitting frequency $\Delta f$. The lines connect the data points as guides to the eye; the SQ rate $\gamma_0$ measured over the same range is shown for comparison and is field independent within the error bars.}
\label{fig:relaxation}
\end{figure}

\section{DQ and SQ relaxation rates}\label{sec:rates}
To obtain a comprehensive understanding of the three-level ground-state dynamics of the $V_{\mathrm{B}}^{-}$ defect (Fig.~\ref{fig:dq-driving}(d)), it is necessary to characterize both the double-quantum relaxation channel between the $\lvert +1 \rangle$ and $\lvert -1 \rangle$ states, described by the rate $\gamma$, and the SQ relaxation channels ($T_1$) between the $\lvert 0 \rangle$ and $\lvert \pm 1 \rangle$ states, described by the rate $\gamma_{\pm}$, which are approximately equal and denoted as $\gamma_0$.
In the following, we experimentally probe these relaxation processes using the pulse sequences shown in Fig.~\ref{fig:relaxation}(a), enabling quantitative extraction of the relaxation rates $\gamma_0$ and $\gamma$. Green laser pulses are used to initialize (polarize) and optically readout the ground-state spin population of the $V_{\mathrm{B}}^{-}$ defects, while microwave $\pi_{0, \pm1}$ pulses are applied to drive transitions between $\lvert 0 \rangle$ and $\lvert \pm 1 \rangle$ states.

Under the assumption of complete relaxation within the three-level ground-state system, the population dynamics of the $V_{\mathrm{B}}^{-}$ defect can be described using the same formalism developed for NV centers in diamond (see Appendix~\ref{app:simulation}). Within this framework, the SQ and DQ relaxation processes contribute differently to the temporal evolution of the spin-state populations. Based on the calculations in Appendix~\ref{app:simulation}, the relaxation rates $\gamma_0$ and $\gamma$ can be quantitatively extracted from the measured population dynamics, which follow $\rho_{0}(\tau) \sim a\, e^{-3\gamma_0 \tau}$ and
$\rho_{\pm 1}(\tau) \sim a'\, e^{-3\gamma_0 \tau} \pm b\, e^{-(\gamma_0 + 2\gamma)\tau}$ \cite{PhysRevLett.118.197201}, where $\tau$ is the evolution time between initialization and readout.
Figure~\ref{fig:relaxation}(b) presents the measured SQ and DQ relaxation processes of the $V_{\mathrm{B}}^{-}$ defects at an external magnetic field of $8.8$~mT. Black squares and red circles denote the SQ and DQ data, respectively. Fits based on the calculated rate-equation model yield relaxation rates of $\gamma_0 = 42.62\pm0.87$~kHz and $\gamma = 46.41\pm0.95$~kHz.
The magnetic-field dependence of the DQ relaxation rate $\gamma$ provides insight into the electric-field and strain noise sampled by the $V_{\mathrm{B}}^{-}$ spin system \cite{PhysRevLett.113.027602}.
As shown in Fig. \ref{fig:relaxation}(c), the observed dependence follows a power-law form
$\gamma = \gamma_{\infty} + A_{\mathrm{eff}}(f_{\mathrm{ref}}/\Delta f)^{\alpha_{\mathrm{eff}}}$,
where $\Delta f$ denotes the ODMR splitting frequency, $A_{\mathrm{eff}}$ is the effective ensemble noise amplitude at the reference frequency $f_{\mathrm{ref}}$, and $\alpha_{\mathrm{eff}}$ characterizes the measured spectral scaling. The $\gamma_{\infty}$ represents a background relaxation rate that is approximately field independent over the measured range. The field-dependent term arises from electric-field and strain noise sampled at the DQ transition frequency.
The measured relaxation rate $\gamma$ decreases with increasing out-of-plane magnetic field, contained implicitly in $\Delta f$, and eventually saturates. The measured data indicate that, within the magnetic-field range explored in our measurements, the field-dependent contribution varies weakly and remains small compared to the field-independent background term $\gamma_{\infty}$.
As analyzed in Appendix~\ref{app:noise}, these data are compatible with correlated-surface and volume-distributed fluctuations, while short-range uncorrelated surface-noise models are disfavored within the assumptions of our ensemble-depth analysis. The present single-sample measurement therefore constrains the noise channel but does not uniquely determine the microscopic origin or spatial localization of the fluctuators.
Over the same magnetic-field range, the SQ relaxation rate $\gamma_0$ shows no significant variation and can be regarded as effectively field independent.
\begin{figure}[!htbp]
  \centering
  \includegraphics[width=\linewidth]{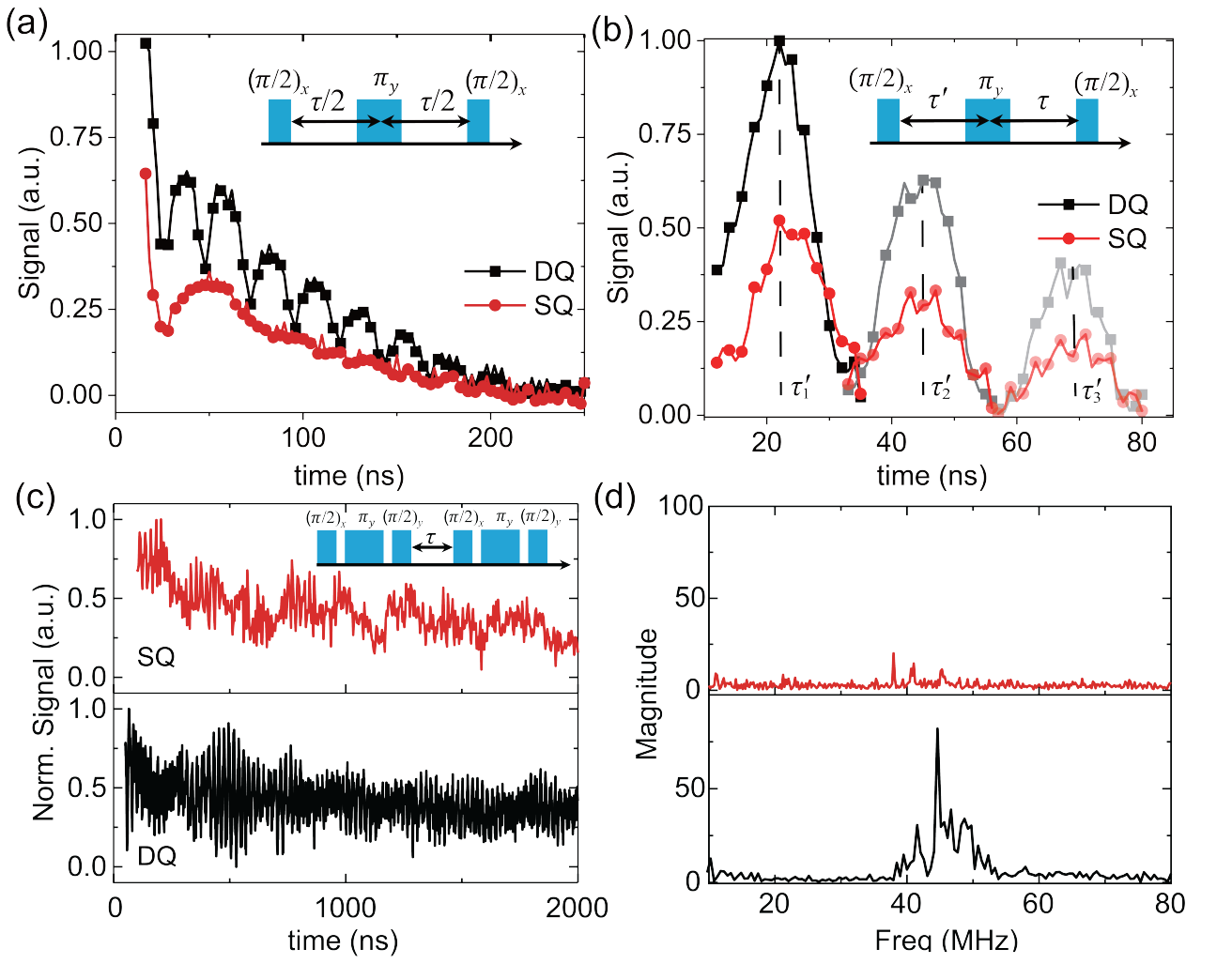}
  \caption{Coherence protection measurements. (a) Spin-echo decay envelopes.
  (b) Representative spin-echo signals acquired at evolution times $\tau' = 22$, $43$, and $64$ ns.
  (c) Spin-echo correlation spectroscopy results.
  (d) The Fast Fourier Transform results of spin-echo correlation spectroscopy. Both red lines are SQ driving, and black lines are DQ driving.}
\label{fig:coherence}
\end{figure}
\section{Spin echo and CPMG correlation}
For the hBN samples used here, the noise contribution from the c-axis electric-field component is weak. This weak electric-field noise is favorable for coherence. Combined with DQ driving and dynamical decoupling protocols to extend its coherence time, the platform could enable AC field sensing.
Based on the measured Rabi oscillations, we choose a Rabi frequency of $41.67$ MHz, corresponding to a $\pi$-pulse duration of $12$ ns, which defines the pulse parameters used in the subsequent spin-echo sequences.
In a spin-echo sequence, the initial $\pi/2$ pulse creates transverse spin coherence, the intermediate $\pi$ pulse reverses the accumulated phase arising from quasi-static dephasing, and the echo formed at $\tau$ reveals the intrinsic coherence limited by dynamical noise.
Increasing the number of intermediate $\pi$ pulses generalizes the spin-echo sequence to the Carr-Purcell-Meiboom-Gill (CPMG) sequence, enabling enhanced suppression of low-frequency noise.
The spin-echo envelope signals under zero magnetic field are depicted in Fig. \ref{fig:coherence}(a), and are well described by a simple exponential decay model, $S(\tau)\propto\exp(-\tau/T_2)$, resulting in the coherence time $T_2$.
Figure \ref{fig:coherence}(b) shows the measured spin-echo signals. Here, the first inter-pulse delay $\tau'$ is fixed, while the second delay
$\tau$ is swept as a control parameter. A pronounced peak-shaped signal is observed centered at $\tau=\tau'$, corresponding to the spin-echo condition. The use of DQ driving leads to an enhanced signal contrast, consistent with the behavior observed in the Rabi oscillations.

Multi-pulse dynamical decoupling protocols, integrated with correlation spectroscopy, enable the phase-coherent accumulation of signals across repeated sequences.
This approach effectively filters out low-frequency noise, thereby enhancing both the sensitivity and frequency selectivity of AC field detection \cite{laraoui2013high}.
In our implementation, correlation spectroscopy utilizes a spin-echo-based sequence, where the evolution time is determined by the temporal separation between two identical spin-echo blocks.
The correlation spectrum obtained in the absence of external radio-frequency (RF) fields is presented in Fig.~\ref{fig:coherence}(c), with the corresponding microwave pulse sequence shown in the inset.
The resulting Fast Fourier Transform (FFT) of this signal (Fig.~\ref{fig:coherence}(d)) exhibits a distinct resonance at approximately 46.3~MHz, and this despite the lack of an external AC field. Indeed, this oscillatory feature originates from the hyperfine interaction between the  electronic ground state of the $V_{\mathrm{B}}^{-}$ defects and the neighboring nuclear spins.
This intrinsic coupling induces a coherent phase accumulation during free evolution intervals, which the spin-echo sequence preserves and encodes into the correlation signal.

\begin{figure}[!htbp]
  \centering
  \includegraphics[width=\linewidth]{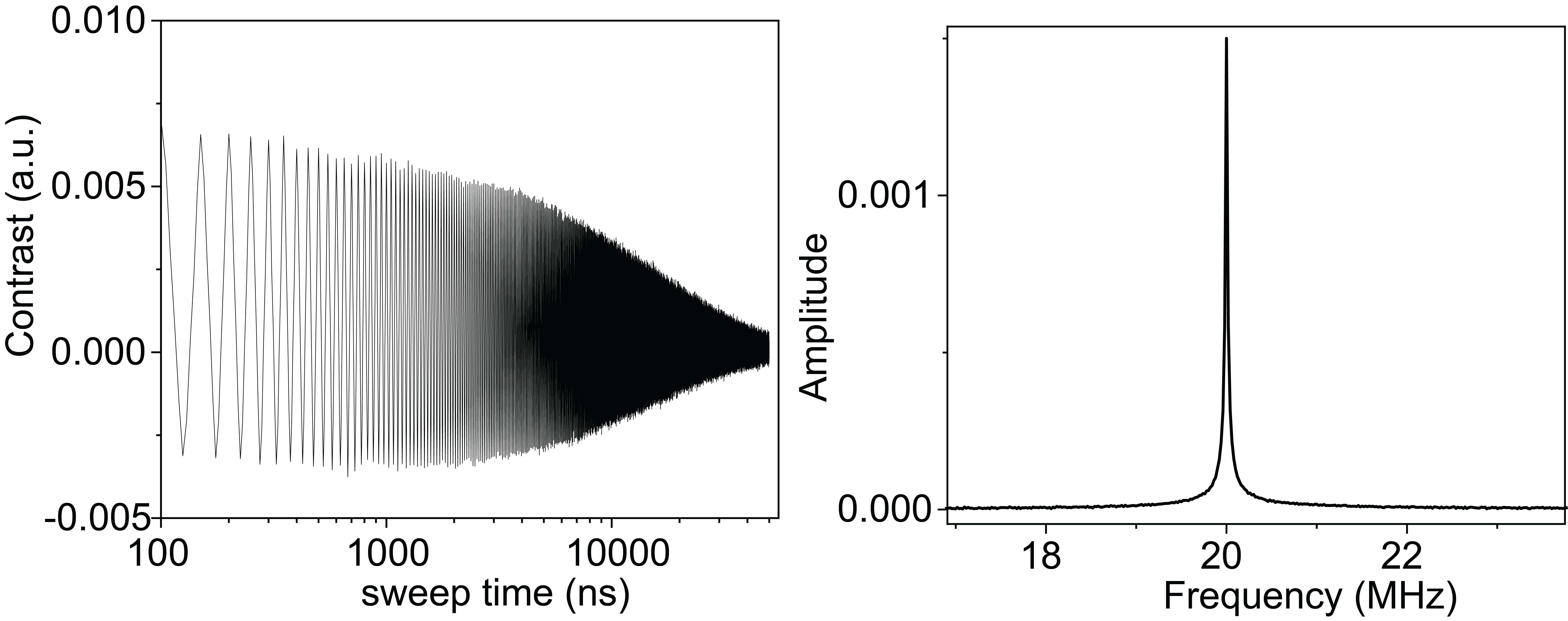}
  \caption{Characterization of external RF signals using spin-echo correlation spectroscopy. (Left) Spin-echo correlation spectrum of the $V_{\mathrm{B}}^{-}$ defect ensemble under exposure to an external RF signal. The oscillation in contrast as a function of sweep time reveals the coherence evolution influenced by the oscillating magnetic field. (Right) FFT of the time-domain correlation signal, showing a sharp peak at 20 MHz, which precisely identifies the frequency of the detected external RF signal.}
\label{fig:rf-correlation}
\end{figure}

An additional RF signal is then applied via a small multi-turn coil placed near the hBN sample, generating a local AC field to drive the boron vacancies. The correlation signal, shown in Fig.~\ref{fig:rf-correlation} (left), oscillates at the frequency of the applied RF field, with a decay trend that remains below the $T_1$ limit, and its Fourier transform in Fig.~\ref{fig:rf-correlation} (right) gives a sharp peak at 20~MHz, confirming that the same protocol detects an externally applied signal.

\section{RF signal-to-noise enhancement in AC magnetometry}
\begin{figure}[!htbp]
  \centering
  \includegraphics[width=\linewidth]{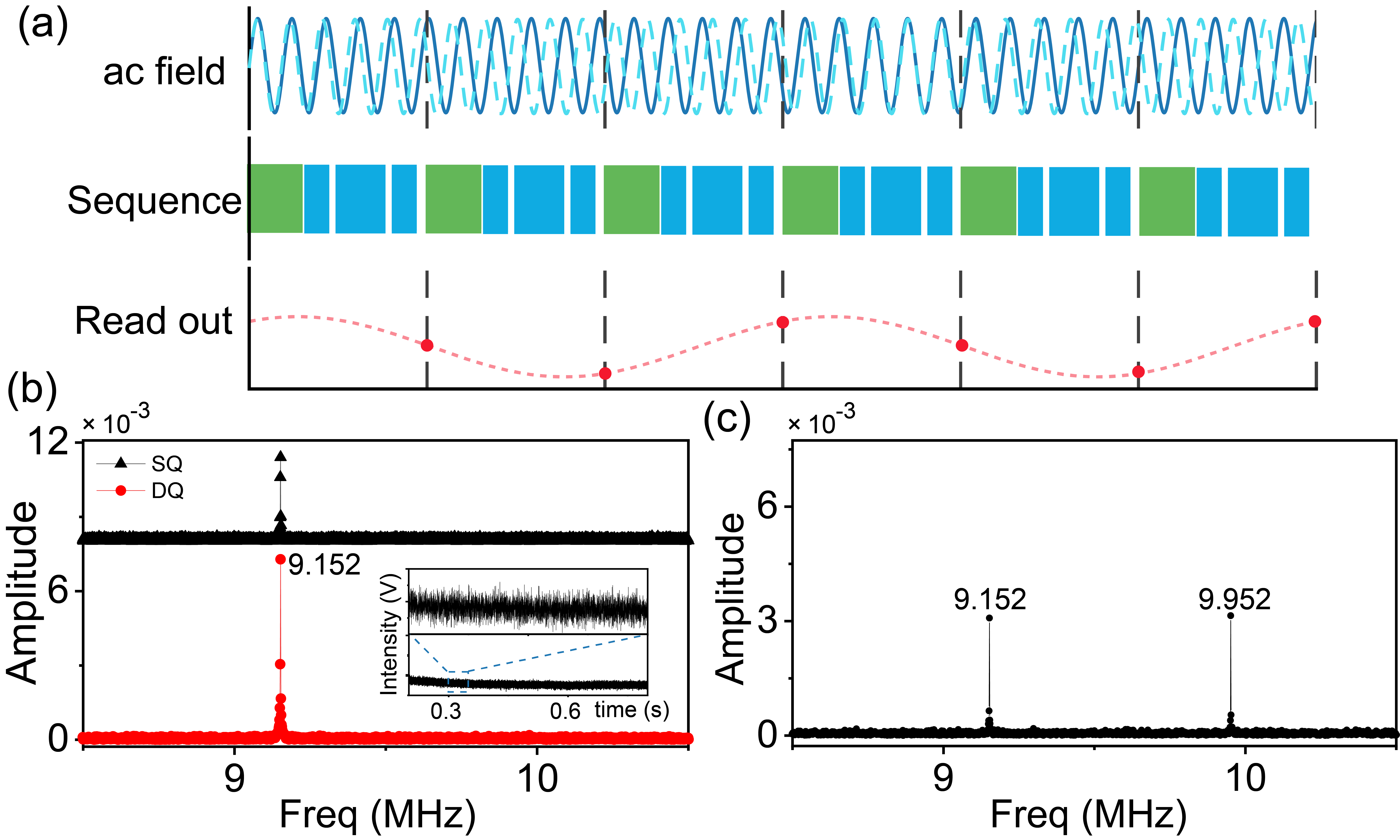}
  \caption{Quantum heterodyne measurement for DQ driving.
  (a) Top: Applied AC signal consisting of two RF fields with closely spaced frequencies, illustrated by two blue traces. Middle: Spin-echo sequence used for phase-coherent detection. Blue rectangles denote microwave pulses, with the inter-pulse delay $\tau$ fixed near $1/(2f_{ac})$. Green rectangles indicate the laser pulses used for optical spin readout.
  Bottom: Measured fluorescence signal (red circles), whose intensity depends on the accumulated spin phase acquired during each spin-echo measurement.
  (b) Comparative fluorescence signals and their FFT spectra. The upper panel shows the signal in the SQ (black triangles) regime, while the lower panel displays the DQ (red circles) regime under a single applied AC field. Inset: The raw time-domain fluorescence intensity.
  (c) FFT spectrum of the fluorescence signal measured under two closely spaced AC field frequencies in the DQ regime. }
\label{fig:qdyne}
\end{figure}

While correlation spectroscopy offers frequency-selective detection of AC fields, its efficiency is fundamentally limited by the requirement to sweep evolution times across multiple pulse sequences to reconstruct the spectrum. To overcome this, we employ quantum heterodyne (Qdyne) detection, which bypasses parameter sweeping by directly sampling the signal phase relative to a stable classical clock \cite{rizzato2023extending}. By executing identical pulse sequences over extended timescales, Qdyne facilitates the coherent accumulation of signal phase, enabling the resolution of AC fields with superior sensitivity and ultra-narrow frequency resolution.
As illustrated in Fig.~\ref{fig:qdyne}(a), we consider an external AC magnetic field defined as $B\cos(2\pi f_{ac} t + \phi)$, where $B$ is the field amplitude, $f_{ac}$ is the frequency, and $\phi$ is the initial phase. The phase $\Phi$ accumulated by the spin state during the $n$-th spin-echo measurement cycle is expressed as:
$\Phi = \frac{2 g \mu_B B}{\pi \hbar f_{ac}}\cos\!\left(2 n \pi \delta T_L + \varphi\right)$.
In this expression, $\delta = f_{ac} - f_{lo}$ represents the frequency detuning, while $T_L$ denotes the total duration of a single sequence block. The term $f_{lo} = i/T_L$ (where $i$ is an integer) defines the central frequency of the synchronized readout protocol. Following the Nyquist-Shannon sampling theorem, the alias-free bandwidth for AC field detection is constrained to $f_{lo} \pm 1/(2T_L)$.

As illustrated in Fig.~\ref{fig:qdyne}(b), and guided by the DQ results from Fig.~\ref{fig:coherence}(a), we applied an external RF field at $20~\mathrm{MHz}$ with a single sequence duration set to $T_L = 9218~\mathrm{ns}$. By selecting the integer parameter $i = 100$, the central frequency of the synchronized readout protocol is established at $f_{lo} = i/T_L = 10.848~\mathrm{MHz}$.
The resulting fluorescence readout for DQ driving and its FFT reveals a measured detuning of $\delta = 9.152~\mathrm{MHz} \pm 1.5~\mathrm{kHz}$, yielding a SNR of 257 (red circles in Fig.~\ref{fig:qdyne}(b)). Consequently, the applied RF frequency is reconstructed as $f_{ac} = f_{lo} + \delta = 20~\mathrm{MHz}$. A benchmark measurement using an SQ spin-echo sequence under identical conditions yields a SNR of 100 (black triangles in Fig.~\ref{fig:qdyne}(b)), giving a 2.57-fold DQ/SQ signal-to-noise improvement without requiring an absolute calibration of the RF coil.
We further demonstrate the versatility of this synchronized readout protocol for multi-frequency sensing, as shown in Fig.~\ref{fig:qdyne}(c). Here, two RF fields were applied simultaneously: a reference signal at frequency $f_0$ and a target signal at an unknown frequency $f_1$.
In the FFT spectrum, the frequency separation between the two peaks directly corresponds to the detuning $\Delta$. By identifying the spectral component of the known tone $f_0$, the unknown frequency is uniquely determined as $f_1 = f_0 \pm \Delta$. The FFT spectrum in Fig.~\ref{fig:qdyne}(c) clearly resolves two components separated by $0.8~\mathrm{MHz}$, confirming the detection of both signals.
This sensing scheme offers significant flexibility in the selection of $T_L$ and $i$. Crucially, because the spectral resolution is independent of the quantum coherence lifetime ($T_2$), it can be continually refined by increasing the total integration time. In practice, the ultimate resolution is governed by the stability of the local oscillator, which dictates the maximum viable $T_L$ and determines the Allan variance of the measurement system.

\section{Conclusion}
In summary, we have demonstrated coherent control of the full spin-triplet ground state of $V_{\mathrm{B}}^{-}$ centers in hBN by implementing a double-quantum (DQ) driving scheme. Compared with conventional SQ methods, the DQ protocol approximately doubles the optical readout contrast in our ensemble and yields a 2.57-fold signal-to-noise improvement in Qdyne AC-field detection under identical conditions. By characterizing the relaxation rates $\gamma_0$ and $\gamma$, we use the DQ channel as a differential probe of electric-field and strain noise. The observed field dependence is compatible with correlated-surface or volume-distributed fluctuations, while short-range uncorrelated surface-noise models are disfavored within our model assumptions. Determining the microscopic origin of this noise will require controlled thickness, density, temperature, or surface-treatment studies. These results establish DQ coherent control and frequency-resolved AC magnetometry in hBN spin ensembles, providing a basis for future optimization of two-dimensional quantum sensors.
\begin{acknowledgments}
We acknowledge support by Singapore National Research Foundation (NRF-CRP30-2023-0003, NRF-CRP31-0001, NRF2023-ITC004-001 and NRF-MSG-2023-0002) and Singapore Ministry of Education Tier 2 Grant (MOE-T2EP50125-0014 and MOE-T2EP50222-0018).
\end{acknowledgments}

\section*{Data availability}
The data that support the findings of this article are available from the corresponding author upon reasonable request.

\appendix

\section{Experimental setup and magnetic field alignment}\label{app:setup}
The sample used in this work is similar to that in our previous work.
The hBN flake, several micrometers thick, was transferred onto a CPW patterned on a 200-nm-thick Au film. Radio-frequency signals, amplified by a microwave amplifier (ZHL-16W-43+), are delivered to the CPW to generate a strong local AC magnetic field at the position of the hBN flake, thereby driving the $\lvert 0 \rangle \leftrightarrow \lvert \pm 1 \rangle$ transitions.
For all ODMR measurements, laser on/off modulation, as well as the TTL trigger signals required by the detection hardware and the AWG, are generated by a pulse generator (Pulse Streamer 8/2).
In addition, a three-dimensional Helmholtz coil (from East Changing Technologies, lnc.) is used to generate the magnetic field, providing each uniaxial static DC magnetic field tunable between $\pm12$ mT.

All spin-control measurements were conducted using a home-built setup. The treated sample is mounted on a custom-designed PCB board and secured on an XYZ manual stage (Newport M-562F-XYZ) and an XYZ piezo scanner (PI P-611.3S), which is controlled by a digital piezo controller (PI E-727.3SD) for precise alignment of $V_{\mathrm{B}}^{-}$ defects with a microscope objective (Nikon CFI60 100X/0.80). Wire bonding was used to connect the CPW and the PCB board \cite{Ruprl2024}.

The experimental apparatus utilizes a 532 nm green laser as the excitation pump source.
To enable precise temporal control of the pump, the laser's amplitude is modulated via an Acousto-Optic Modulator (AOM). This modulation is driven by a TTL signal sourced from a digital pulse sequencer (Swabian Instruments Pulse Streamer 8/2), allowing for high-resolution pulse-width and repetition-rate control.
The resulting photoluminescence (PL) is isolated using a Thorlabs DMLP567R dichroic mirror and a Thorlabs FELH0750 long-pass filter, then coupled into a $\varnothing$50 µm, 0.22 NA multi-mode fiber and routed to an amplified photodiode for detection.

All digital control and timing for the spin manipulation sequences are executed by the Pulse Streamer 8/2, which serves as the central orchestration unit. This device also synthesizes both reference signals required for lock-in demodulation to ensure synchronous operation.
Specifically, the period of reference signal 1 is precisely matched to the duration of the experimental timing sequence.
Signal acquisition is performed with a lock-in time constant of 300 ms, with the output subsequently averaged over a 1-second buffer to optimize the signal-to-noise ratio.
For quantum heterodyne measurements, a DAQ-6343 is employed, with its readout gating window strictly controlled by the Pulse Streamer to maintain phase-coherent acquisition.
The entire experimental suite—spanning real-space mapping to complex time-domain pulse sequences—is fully automated via a custom Python-based data acquisition framework.

\section{Microwave control for double quantum driving}\label{app:microwave}
To achieve high-fidelity coherent control in double quantum driving, we employ a vector modulation scheme using a Stanford Research Systems SG384 Signal Generator (equipped with an External I/Q modulation option) and a Tektronix AWG5208 Arbitrary Waveform Generator. This setup allows for precise manipulation of the microwave field’s frequency, phase, and amplitude by defining the relationship between the In-phase ($I$) and Quadrature ($Q$) baseband signals.

\subsection{Mathematical principles of vector modulation}
The microwave output $S_{RF}(t)$ is synthesized by mixing the baseband inputs from the AWG with a carrier signal at frequency $\omega_c$:
\begin{equation}
    S_{RF}(t)=I(t)\cos(\omega_ct)-Q(t)\sin(\omega_ct).
\end{equation}
By modulating $I(t)$ and $Q(t)$ as harmonic functions of a target detuning frequency $\omega_m$, we can selectively generate specific spectral components required for the driving protocol:
\begin{itemize}
    \item Double Sideband (DSB): Driving $I(t) = \cos(\omega_m t)$ while setting $Q(t) = 0$ yields a suppressed-carrier signal with power split between two sidebands:\begin{equation}
        S_{RF}(t) = \frac{1}{2} \left[ \cos((\omega_c + \omega_m)t) + \cos((\omega_c - \omega_m)t) \right].
    \end{equation}
    \item Blue Sideband (USB): Setting $I(t) = \cos(\omega_m t)$ and $Q(t) = \sin(\omega_m t)$ constructively interferes at the upper sideband and cancels at the lower:
    \begin{equation}
        S_{RF}(t) = \cos((\omega_c + \omega_m)t).
    \end{equation}
    \item Red Sideband (LSB): Setting $I(t) = \cos(\omega_m t)$ and $Q(t) = -\sin(\omega_m t)$ yields the lower sideband:\begin{equation}
        S_{RF}(t) = \cos((\omega_c - \omega_m)t).
    \end{equation}
\end{itemize}

\subsection{Experimental configuration and synchronization}
The precision of this phase-coherent control relies on rigid timing synchronization across the instrument stack. All devices are slaved to a common 10 MHz reference clock to eliminate relative frequency drift.
In our architecture, the master time-base is provided by the Zurich Instruments MFLI (500 kHz Lock-in Amplifier). This 10 MHz reference is distributed to
\begin{itemize}
    \item[1.] Tektronix AWG5208: Ensures phase-stable generation of the $I$ and $Q$ baseband envelopes.
    \item[2.] SRS SG384: Synchronizes the microwave carrier, maintaining a fixed phase relationship with the modulation signals.
    \item[3.] Pulse Streamer 8/2: Aligns the digital pulse sequences (TTL) for experimental timing and instrument triggering, ensuring that the microwave gates are perfectly coincident with the modulated waveforms.
\end{itemize}
By integrating the Pulse Streamer into the same clock domain, we can achieve high timing precision between the digital "ON/OFF" gating and the microwave phase. This is vital for coherent protocols where the relative phase of the driving field at the moment of pulse arrival must be strictly deterministic.

\section{Population dynamics simulation of hBN three-level systems}\label{app:simulation}

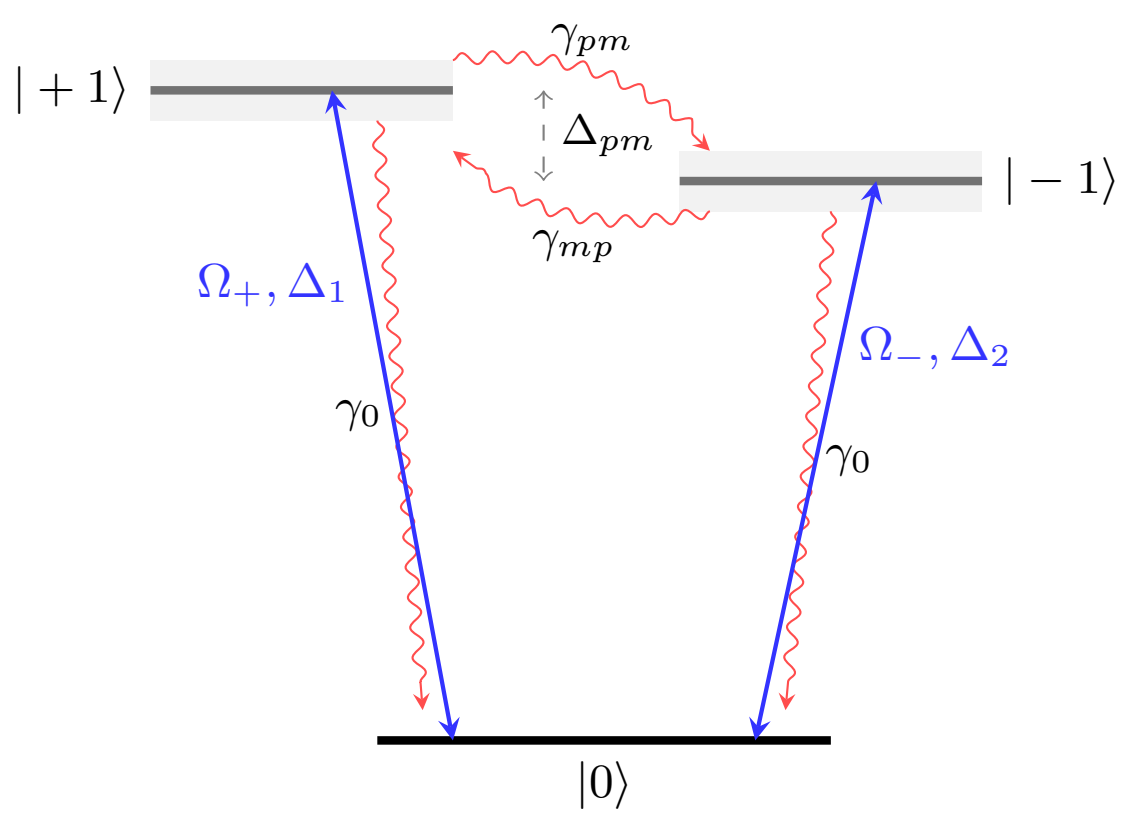
\begin{figure}[!htbp]
\centering
\begin{tikzpicture}[
  level/.style={ultra thick, black},
  decay/.style={->, >=stealth, decorate, decoration={snake, amplitude=0.4mm, segment length=2mm, post length=1mm}, red!70},
  drive/.style={<->, >=stealth, thick, blue!80},
  broadening/.style={fill=gray!20, opacity=0.5},
  font=\small
]
\draw[level] (0.5,0) -- (3.5,0) node[below, pos=0.5] {$|0\rangle$}; 

\draw[level] (-1,4.3) node[left] {$|+1\rangle$} -- (1,4.3); 

\draw[level] (2.5,3.7) -- (4.5,3.7) node[right] {$|-1\rangle$};

\fill[broadening] (-1,4.1) rectangle (1,4.5);
\fill[broadening] (2.5,3.5) rectangle (4.5,3.9);

\draw[<->, dashed, gray] (1.6, 4.3) -- node[right, black, font=\footnotesize] {$\Delta_{pm}$} (1.6, 3.7);

\draw[decay] (0.5, 4.1) -- node[left, black] {$\gamma_{0}$} (0.8, 0.2);
\draw[decay] (3.5, 3.5) -- node[right, black] {$\gamma_{0}$} (3.2, 0.2);

\draw[drive] (1.0, 0) -- node[left, pos=0.7] {$\Omega_+, \Delta_1$} (0.2, 4.3);
\draw[drive] (3.0, 0) -- node[right, pos=0.7] {$\Omega_-, \Delta_2$} (3.8, 3.7);

\draw[decay] (1., 4.5) to[bend left=25] node[above, black] {$\gamma_{pm}$} (2.7, 3.9);
\draw[decay] (2.7, 3.5) to[bend left=25] node[below, black] {$\gamma_{mp}$} (1, 3.9);

\end{tikzpicture}
\caption{Schematic of the three-level spin system model for the $V_{\mathrm{B}}^{-}$ defect in hBN.}
\label{fig:spin-model}
\end{figure}

The model used to capture the population dynamics is shown in {{Fig. \ref{fig:spin-model}}}. The system consists of three states driven by one or two microwave excitations and incoherent scattering processes.

We model the coupled coherent and incoherent dynamics using the Lindblad master equation:
\begin{equation}
\frac{d\rho}{dt} = -i[H(t), \rho] + \sum_k \left( L_k \rho L_k^\dagger - \frac{1}{2} \{L_k^\dagger L_k, \rho\} \right),
\end{equation}
where the Hamiltonian $H(t)$ governs the coherent evolution and the set of collapse operators $\{L_k\}$ accounts for dissipative processes.

Throughout this analysis, we adopt the following \textbf{basis order}:
\begin{equation}
\{ |+1\rangle, |0\rangle, |-1\rangle \}.
\end{equation}
The matrix element $\rho_{00}$ in the numerical implementation corresponds to the population of the ground state, $P_0 = \langle 0|\rho|0\rangle$.

\subsection{Hamiltonian and collapse operators}

For the bare Hamiltonian $H_0$, we define two transition frequencies (in Hz):
\begin{equation}
f_{0+} = f_{\text{avg}} + \frac{\Delta_{pm}}{2} + \delta f_{+}, \qquad
f_{0-} = f_{\text{avg}} - \frac{\Delta_{pm}}{2} + \delta f_{-},
\end{equation}
where $\Delta_{pm}$ represents the measured splitting between the $|+1\rangle$ and $|-1\rangle$ states, and $\delta f_{\pm}$ denote static frequency offsets incorporated for ensemble modeling.

In the laboratory frame (without the Rotating Wave Approximation, RWA), the bare Hamiltonian is expressed as:
\begin{equation}
H_{0}^{\text{(full)}} = \hbar \operatorname{diag}(\omega_{0+}, \, 0, \, \omega_{0-}), \qquad \omega_{0\pm} = 2\pi f_{0\pm}.
\end{equation}

In the \textbf{average rotating frame}, we rotate at the reference frequency $f_{\text{avg}}$ to yield a symmetric splitting Hamiltonian:
\begin{equation}
H_{0}^{\text{(RWA)}} = \hbar \operatorname{diag}\left( +\frac{\omega_{\Delta}}{2}, \, 0, \, -\frac{\omega_{\Delta}}{2} \right), \qquad \omega_{\Delta} \approx 2\pi \Delta_{pm}.
\end{equation}
To account for ensemble realizations, we include static level shifts $\delta\omega_{+}$ and $\delta\omega_{-}$, modifying the diagonal terms:
\begin{equation}
H_0^{\text{(RWA)}} \to \hbar \operatorname{diag}\left( +\frac{\omega_\Delta}{2} + \delta\omega_{+}, \, 0, \, -\frac{\omega_\Delta}{2} + \delta\omega_{-} \right).
\end{equation}

The transition operators are defined as:
\begin{equation}
\begin{aligned}
S_{+} &= |+1\rangle\langle 0|, \quad &S_{-} &= |-1\rangle\langle 0|, \\
S_{+}^\dagger &= |0\rangle\langle +1|, \quad &S_{-}^\dagger &= |0\rangle\langle -1|.
\end{aligned}
\end{equation}

\paragraph{Drive Hamiltonian and Cross-Driving}
We consider one or two \textbf{phase-locked} microwave tones targeting the $|0\rangle \leftrightarrow |+1\rangle$ and $|0\rangle \leftrightarrow |-1\rangle$ transitions. A central feature of this model is \textbf{cross-driving}, where a tone resonant with one transition simultaneously drives the other off-resonantly.

In the laboratory frame, each tone couples via the operator
\begin{equation}
V \equiv S_{+} + S_{-} + S_{+}^\dagger + S_{-}^\dagger.
\end{equation}
For a tone at frequency $\omega_d = 2\pi f_d$ with an angular Rabi rate $\Omega_d$, the drive Hamiltonian is:
\begin{equation}
H_D^{\text{(full)}}(t) = \hbar \Omega_d \cos(\omega_d t) V.
\end{equation}

The specific experimental configurations are modeled as:
\begin{itemize}
    \item {SQ EXP1:} Single-quantum drive on the $|0\rangle \leftrightarrow |+1\rangle$ transition,
    $H_D(t) = \hbar \Omega_{+} \cos(\omega_{0+} t) V$.

    \item {SQ EXP2:} Single-quantum drive on the $|0\rangle \leftrightarrow |-1\rangle$ transition,
    $H_D(t) = \hbar \Omega_{-} \cos(\omega_{0-} t) V$.

    \item {DQ EXP3:} Double-quantum drive applying both tones simultaneously,
    $H_D(t) = \hbar \Omega_{+} \cos(\omega_{0+} t) V + \hbar \Omega_{-} \cos(\omega_{0-} t) V$.
\end{itemize}

\paragraph{General RWA in the Average Rotating Frame}
We define the rotating frame relative to a \textbf{fixed} reference frequency $\omega_{\text{avg}} = 2\pi f_{\text{avg}}$, which is independent of the specific tone resonance. In this formulation, we do \textbf{not} a priori assign a specific tone to a specific transition; instead, each tone is allowed to couple to all available transitions, and the appropriate detunings emerge naturally.

After applying the RWA (neglecting terms oscillating at $\pm(\omega_k + \omega_{\text{avg}})$), the general two-tone drive Hamiltonian in the $\omega_{\text{avg}}$ rotating frame is:
\begin{equation}
\begin{split}
H_D^{\text{(RWA)}}(t) = \frac{\hbar}{2} \sum_{k \in \{1,2\}} \Omega_k \big[ & e^{-i\Delta_k t} (S_{+} + S_{-}) \\
& + e^{+i\Delta_k t} (S_{+}^\dagger + S_{-}^\dagger) \big],
\end{split}
\end{equation}
where $\Delta_k \equiv \omega_k - \omega_{\text{avg}}$. Setting $\Delta_1 = +\omega_\Delta/2$ and $\Delta_2 = -\omega_\Delta/2$ for our specific frequencies yields:
\begin{equation}
\begin{aligned}
H_D^{\text{(RWA)}}(t) = \frac{\hbar}{2} \Big[
&\Omega_1 \Big( e^{-i{\omega_\Delta}t/2}(S_{+} + S_{-}) \\
&\quad + e^{+i{\omega_\Delta}t/2}(S_{+}^\dagger + S_{-}^\dagger) \Big) \\
+ &\Omega_2 \Big( e^{+i{\omega_\Delta}t/2}(S_{+} + S_{-}) \\
&\quad + e^{-i{\omega_\Delta}t/2}(S_{+}^\dagger + S_{-}^\dagger) \Big)
\Big].
\end{aligned}
\end{equation}
This general form is particularly robust for ensemble broadening simulations where transition frequencies may vary relative to the drive.

\paragraph{Incoherent Processes}
Dissipation in this system is characterized by symmetric exchange rather than unidirectional decay:
\begin{itemize}
    \item \textbf{Spin-environment scattering:} Transitions $|0\rangle \leftrightarrow |\pm 1\rangle$ occur via symmetric scattering at a rate $\gamma_0$.
    \item \textbf{Excited-state mixing:} Transitions between $|+1\rangle \leftrightarrow |-1\rangle$ are modeled with rates $\gamma_{pm}$ and $\gamma_{mp}$.
    \item \textbf{Pure dephasing:} Dephasing of the $|\pm 1\rangle$ states is included with rate $\gamma_\phi$.
\end{itemize}

The corresponding collapse operators are:
\begin{equation}
\begin{aligned}
L_{+0} &= \sqrt{\gamma_0} \, |0\rangle\langle +1|, & L_{0+} &= \sqrt{\gamma_0} \, |+1\rangle\langle 0|, \\
L_{-0} &= \sqrt{\gamma_0} \, |0\rangle\langle -1|,& L_{0-} &= \sqrt{\gamma_0} \, |-1\rangle\langle 0|, \\
L_{+-} &= \sqrt{\gamma_{pm}} \, |-1\rangle\langle +1|, & L_{-+} &= \sqrt{\gamma_{mp}} \, |+1\rangle\langle -1|, \\
L_{\phi,+} &= \sqrt{\gamma_\phi} \, |+1\rangle\langle +1|,& L_{\phi,-} &= \sqrt{\gamma_\phi} \, |-1\rangle\langle -1|.
\end{aligned}
\end{equation}

\subsection{Rabi oscillation and incoherent decay analysis}
The simulation proceeds in four stages: coherent evolution, signal projection, ensemble averaging, and parameter optimization.

\begin{figure}[!htbp]
  \centering
  \includegraphics[width=\linewidth]{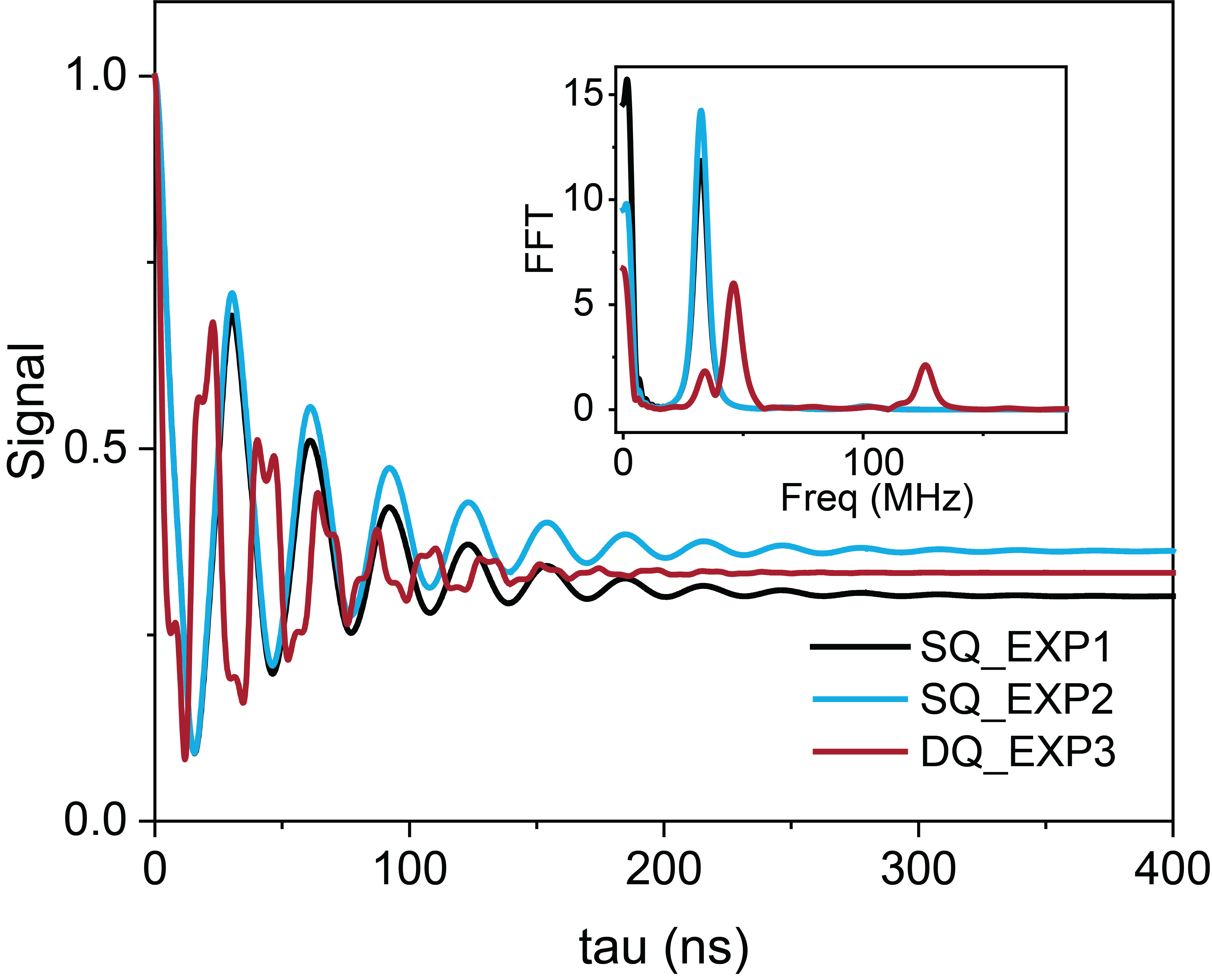}
  \caption{Simulated Rabi results for single $V_{\mathrm{B}}^{-}$ defect, and the inset is its FFT results.}\label{fig:single-spin-rabi}
\end{figure}

First, the system is initialized in the state $|0\rangle$. For a given pulse duration $\tau$, the density matrix $\rho(t)$ evolves according to the Lindblad master equation until $t = \tau$. To connect the microscopic state to macroscopic observations, the experimental readout signal $S(\tau)$ is modeled as a linear combination of the excited-state populations:
\begin{equation}
S(\tau) = A \big[ S_0 - \left( c_{+} P_{+}(\tau) + c_{-} P_{-}(\tau) \right) \big],
\end{equation}
where $P_{\pm}(\tau) = \langle \pm 1 | \rho(\tau) | \pm 1 \rangle$ are the populations of the $|\pm 1\rangle$ levels. The coefficients $A, S_0,$ and $c_{\pm}$ ($0 \le c_{\pm} \le 1$) are phenomenological parameters characterizing the readout contrast and efficiency. The characteristic frequencies of the resulting quantum beats are subsequently analyzed by performing an FFT on the time-domain signal $S(\tau)$.
To validate our model, we first simulate the Rabi oscillations for a single $V_{\mathrm{B}}^{-}$ defect, as presented in Fig.~\ref{fig:single-spin-rabi}.
The black and blue traces represent the population dynamics under SQ driving for the $|0\rangle \leftrightarrow |+1\rangle$ and $|0\rangle \leftrightarrow |-1\rangle$ transitions, respectively, while the red trace depicts the response under DQ driving.
A spectral analysis of these oscillations, shown in the inset of Fig.~\ref{fig:single-spin-rabi}, reveals that the DQ driving frequency is approximately $\sqrt{2}$ times faster than that of the SQ transitions. This frequency enhancement is consistent with the coherent evolution of a three-level system, where the effective coupling strength is amplified when both transitions are driven simultaneously.

For such single-spin models, SQ transitions can achieve very high contrast, especially when the system is under resonant excitation. However, it often fails to capture the spectral complexities observed in experiments.
To resolve these discrepancies, we extend the framework to an ensemble model that accounts for the inhomogeneous broadening. This is implemented by assigning \textbf{independent static frequency shifts}, $\delta f_{+}$ and $\delta f_{-}$, to the $|+1\rangle$ and $|-1\rangle$ levels, respectively, sampled from Gaussian distributions:
\begin{equation}
\delta f_{+} \sim \mathcal{N}(0, \sigma^2), \qquad \delta f_{-} \sim \mathcal{N}(0, \sigma^2).
\end{equation}
The final observable is obtained by simulating $N$ independent realizations and computing the ensemble average:
\begin{equation}
\langle S(\tau) \rangle = \frac{1}{N} \sum_{i=1}^N S_i(\tau).
\end{equation}

\begin{figure}[!htbp]
  \centering
  \includegraphics[width=\linewidth]{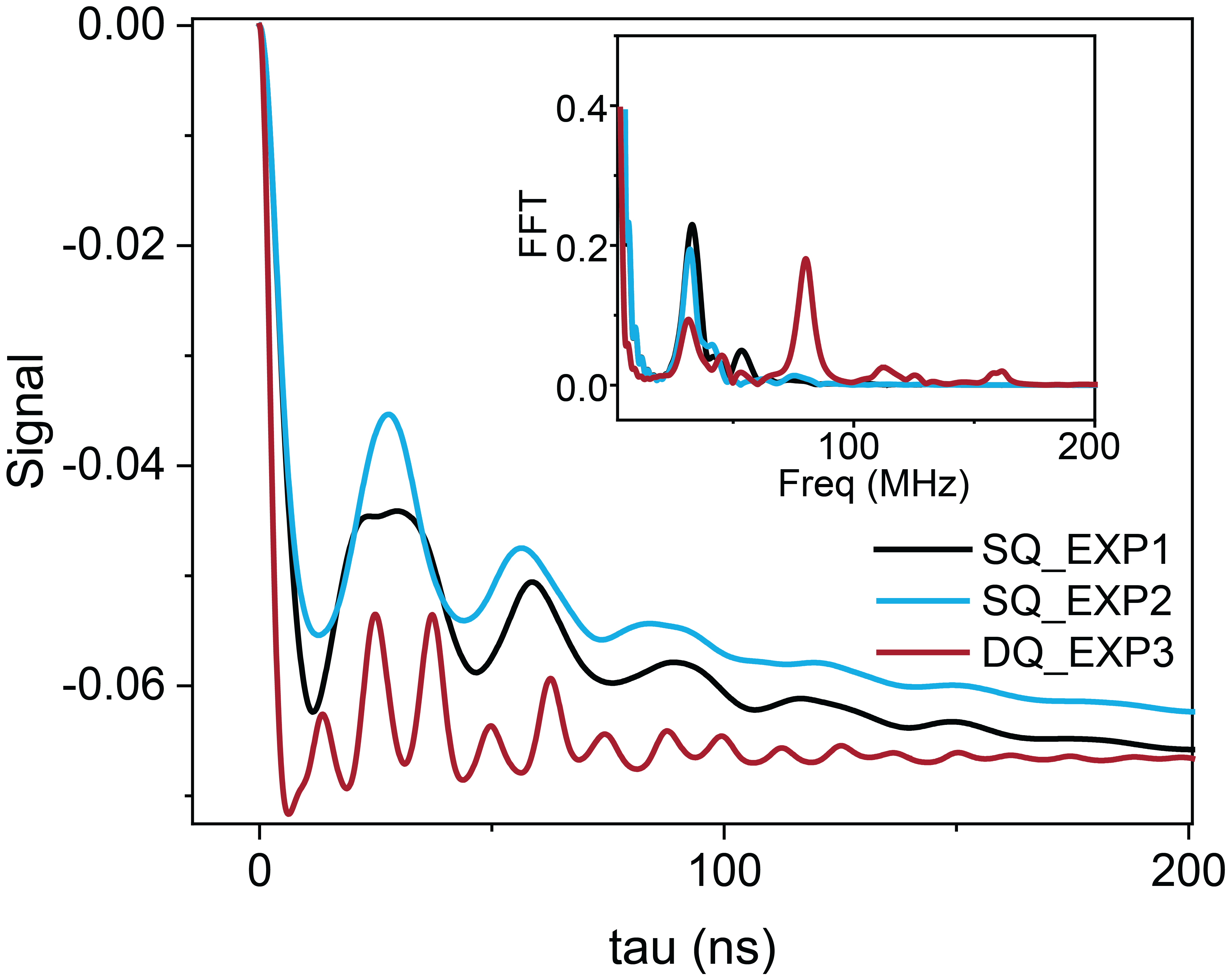}
  \caption{Simulation of Rabi oscillations in an ensemble system composed of $N=81$ independent $V_{\mathrm{B}}^{-}$ defects. To account for inhomogeneous broadening, each of the 81 defects is assigned static frequency shifts $\delta f_{\pm}$ sampled from independent Gaussian distributions $\mathcal{N}(0, \sigma^2)$. The main panel displays the ensemble-averaged time-domain signal, where dephasing results in a damping of the oscillations. The inset displays the corresponding FFT spectra.}\label{fig:ensemble-rabi}
\end{figure}
To investigate the dynamics within an ensemble environment, we constructed a discrete model comprising $N=81$ independent spin defects. Inhomogeneous broadening is incorporated by assigning independent static frequency shifts, $\delta f_{\pm} \sim \mathcal{N}(0, \sigma^2)$, to each defect realization. As shown in Fig. \ref{fig:ensemble-rabi}, the black and blue traces represent the ensemble-averaged Rabi oscillations for the two SQ transitions, while the red trace depicts the response under DQ driving.

Finally, to bridge the gap between theory and experiment, we implement an optimization routine to fit the measured SQ and DQ data. By tuning the collective set of physical and phenomenological parameters $\boldsymbol{\theta} = (\gamma_\phi, \sigma, c_+, c_-, A)$, we minimize the residuals between the simulated ensemble average $\langle S(\tau) \rangle$ and the experimental measurements, ensuring a robust determination of the system's decoherence properties.

\subsection{Analytical solutions of incoherent dynamics}

To isolate the incoherent dynamics of the spin system, we consider the regime where $H = 0$. Under this condition, the evolution is governed solely by the diagonal elements of the Lindblad master equation, which describe the population transfer between states. We define two specific observables to evaluate the relaxation processes:
\begin{itemize}
    \item \textbf{Configuration (a):} The system is initialized in the $|0\rangle$ state ($p_0(0)=1$), and we track the difference $D_b(t) = P_0(t) - P_{-1}(t)$.
    \item \textbf{Configuration (b):} The system is initialized in the $|\pm 1\rangle$ states to track the excited-state population asymmetry, defined as $D_{c,+}(t) = P_{+1}(t) - P_{-1}(t)$ for an initial $|+1\rangle$ state, and $D_{c,-}(t) = P_{-1}(t) - P_{+1}(t)$ for an initial $|-1\rangle$ state.
\end{itemize}
These dynamics are analyzed across multiple temporal scales, ranging from a short window (0--400~ns) to a long-term relaxation window (0--15~$\mu$s).

With the coherent terms removed, the population vector $\mathbf{p} = [p_{+}(t), p_{0}(t), p_{-}(t)]^T$ evolves according to the classical rate equation:
\begin{equation}
\begin{aligned}
\frac{d\mathbf{p}}{dt} &= M \mathbf{p}, \\
M &= \begin{pmatrix}
-(\gamma_0+\gamma_{pm}) & \gamma_0 & \gamma_{mp} \\
\gamma_0 & -2\gamma_0 & \gamma_0 \\
\gamma_{pm} & \gamma_0 & -(\gamma_0+\gamma_{mp})
\end{pmatrix},
\end{aligned}
\end{equation}
where $\gamma_0$ represents the symmetric scattering rate between $|0\rangle$ and $|\pm 1\rangle$, while $\gamma_{pm}$ and $\gamma_{mp}$ characterize the mixing between the $|+1\rangle$ and $|-1\rangle$ states. The general solution is $\mathbf{p}(t) = e^{Mt}\mathbf{p}(0)$. Given that $M$ is a $3 \times 3$ matrix with one zero eigenvalue (representing the steady state), any linear observable $D(t) = \mathbf{v}^T\mathbf{p}(t)$ can be expressed as a sum of at most two decaying exponentials plus a constant:
\begin{equation}
D(t) = D_\infty + a_1 e^{\lambda_1 t} + a_2 e^{\lambda_2 t}.
\end{equation}
The relaxation rates correspond to the non-zero eigenvalues of $M$, specifically $\lambda_1 = -3\gamma_0$ and $\lambda_2 = -(\gamma_0 + \gamma_{pm} + \gamma_{mp})$, which we use as the fitting function in the main text.

\section{Quantitative analysis of the DQ relaxation channels}\label{app:noise}

\subsection{Which noise drives the double-quantum channel}

The DQ transition $\lvert +1 \rangle \leftrightarrow \lvert -1 \rangle$ carries $\Delta m_s = 2$ and is therefore not driven to first order by the transverse magnetic noise responsible for the SQ channels near $D$ \cite{PhysRevLett.118.197201}. It couples instead to transverse electric-field and strain noise, which enters the ground-state spin Hamiltonian as
\begin{equation}
  H_E = \Pi_z S_z^2 + \Pi_x\left(S_y^2 - S_x^2\right)
        + \Pi_y\left(S_x S_y + S_y S_x\right),
  \label{eq:HE}
\end{equation}
where the $\Pi_i$ denote the combined electric-field and strain couplings, and the transverse components $\Pi_x$ and $\Pi_y$ connect $\lvert +1 \rangle$ and $\lvert -1 \rangle$ directly. Fermi's golden rule makes the DQ rate proportional to the transverse noise power spectral density evaluated at $\Delta f = f_+ - f_-$, the ODMR splitting measured in Fig.~\ref{fig:relaxation}(c). The pair $(\gamma_0,\gamma)$ therefore provides differential sensitivity to the magnetic and electric-field/strain channels, but does not by itself identify the spatial location of the fluctuators.

We write a local candidate model as
\begin{equation}
  \gamma(\Delta f, z) = \gamma_{\infty}
    + A_{\mathrm{s}}\left(f_{\mathrm{ref}}/\Delta f\right)^{\alpha} K(z,t),
  \label{eq:app-model}
\end{equation}
where $\gamma_{\infty}$ is the background that is field independent over the measured range, $A_{\mathrm{s}}$ is a local noise scale at $f_{\mathrm{ref}}$, and $K(z,t)$ describes the spatial dependence in a flake of thickness $t$. The amplitude $A_{\mathrm{eff}}$ in Sec.~\ref{sec:rates} is an effective ensemble parameter and should not be identified with $A_{\mathrm{s}}$ by a simple depth average.

\subsection{Competing depth dependences}\label{sec:noise-kernels}

We use three limiting spatial models. For an uncorrelated surface, $K_1(z)=(z_c/z)^p$, with representative exponents $p=2$ and $p=4$ and a cutoff $z_c=0.333$~nm equal to one hBN layer. For a surface correlated over a lateral scale $L$, we use the phenomenological interpolation $K_1(z)=[L^2/(L^2+z^2)]^{p/2}$, which is nearly constant for $z\ll L$ and approaches $(L/z)^p$ for $z\gg L$. The relation between a microscopic charge process and the depth exponent generally depends on its spatial correlation function, so these kernels should not be read as a unique charge-versus-dipole identification \cite{Candido2024surface}. For either surface model, $K(z,t)=K_1(z)+K_1(t-z)$. The volume-distributed null model instead has $K(z,t)=1$.

The sample used here was prepared by 250-keV proton irradiation \cite{Ruprl2024}. We therefore retain a normalized defect profile $P(z)$ explicitly. A uniform profile, $P(z)=1/t$, is used below only as a sensitivity-test baseline; for this special case, the mean uncorrelated kernel reduces for $t\gg z_c$ to
\begin{equation}
  \langle K \rangle (t) \simeq \frac{2 z_c}{(p-1)\,t}.
  \label{eq:kmean}
\end{equation}
This arithmetic mean is dominated by the near-surface cutoff and is not the rate returned by the experiment.

\subsection{Observable matched to the experimental fit}

The experimental three-level analysis gives two relaxation modes,
\begin{equation}
 S_{\mathrm{DQ}}(\tau)=c+a e^{-3\gamma_0\tau}
 \pm b e^{-(\gamma_0+2\gamma)\tau}.
 \label{eq:dq-fit}
\end{equation}
If $\gamma$ depends on depth, the second mode must first be averaged over the ensemble,
\begin{equation}
 C_{\mathrm{DQ}}(\tau,\Delta f)=\int_0^t P(z)
 e^{-[\gamma_0+2\gamma(\Delta f,z)]\tau}\,\mathrm{d}z.
 \label{eq:dq-average}
\end{equation}
We then fit $C_{\mathrm{DQ}}$ with $b' e^{-(\gamma_0+2\gamma_{\mathrm{fit}})\tau}$, which is the same $\gamma$-dependent mode used in Eq.~\eqref{eq:dq-fit}. The simulation uses logarithmically sampled delays from 10~ns to 100~$\mu$s, matching the range plotted in Fig.~\ref{fig:relaxation}(b). This procedure also accounts for a steep surface kernel: defects that relax before the first delay reduce the fitted amplitude rather than producing an arbitrarily large measured rate.

\subsection{Comparison with the measured field dependence}

\begin{table*}[t]
\caption{Outcome of fitting each spatial model to all central DQ rates in Fig.~\ref{fig:relaxation}(c), using the observable in Eq.~\eqref{eq:dq-average}. The comparison assumes a uniform depth profile and therefore tests model sensitivity rather than assigning a unique microscopic source.}
\label{tab:calib}
\begin{ruledtabular}
\begin{tabular}{ll}
Model & Outcome \\
\colrule
uncorrelated, $p = 4$             & $A_{\mathrm{s}}=1.03\times10^{13}$ kHz, $\alpha=5.37$, RMS = 0.73 \\
uncorrelated, $p = 2$             & $A_{\mathrm{s}}=9.83\times10^{6}$ kHz, $\alpha=2.78$, RMS = 0.72 \\
correlated, $L = 0.5~\mu$m        & $A_{\mathrm{s}}=49.1$ kHz, $\alpha=1.57$, RMS = 0.69 \\
correlated, $L = 2~\mu$m          & $A_{\mathrm{s}}=11.1$ kHz, $\alpha=1.46$, RMS = 0.71 \\
correlated, $L = 10~\mu$m         & $A_{\mathrm{s}}=4.58$ kHz, $\alpha=1.46$, RMS = 0.71 \\
volume distributed                & $A_{\mathrm{s}}=8.30$ kHz, $\alpha=1.46$, RMS = 0.71 \\
\end{tabular}
\end{ruledtabular}
\end{table*}

The DQ rate rises from approximately $46$~kHz at large splitting to approximately $55$~kHz at $\Delta f=90$~MHz. For each spatial model, we jointly fit $A_{\mathrm{s}}$ and the local spectral exponent $\alpha$ to all central values in Fig.~\ref{fig:relaxation}(c), while fixing $\gamma_{\infty}=46$~kHz. Under the uniform-profile baseline, the correlated models and the volume model reproduce the curve with $\alpha=1.46$--$1.57$ and local rate scales below $50$~kHz. The uncorrelated $p=2$ and $p=4$ kernels reach comparable residuals only with local scales of order $10^7$ and $10^{13}$~kHz, respectively, together with steeper spectra. These values strongly disfavor short-range uncorrelated surface noise within this model, but they do not constitute a mathematical exclusion or a unique identification of the remaining mechanisms.

In particular, the present measurement does not determine $L$: a correlated surface with $L$ larger than the flake approaches the volume model over the measured thickness. The data therefore support the electric-field/strain assignment from the DQ selection rule, while the spatial origin remains unresolved.

\subsection{A thickness series would discriminate}

\begin{figure*}[!htbp]
  \centering
  \includegraphics[width=\textwidth]{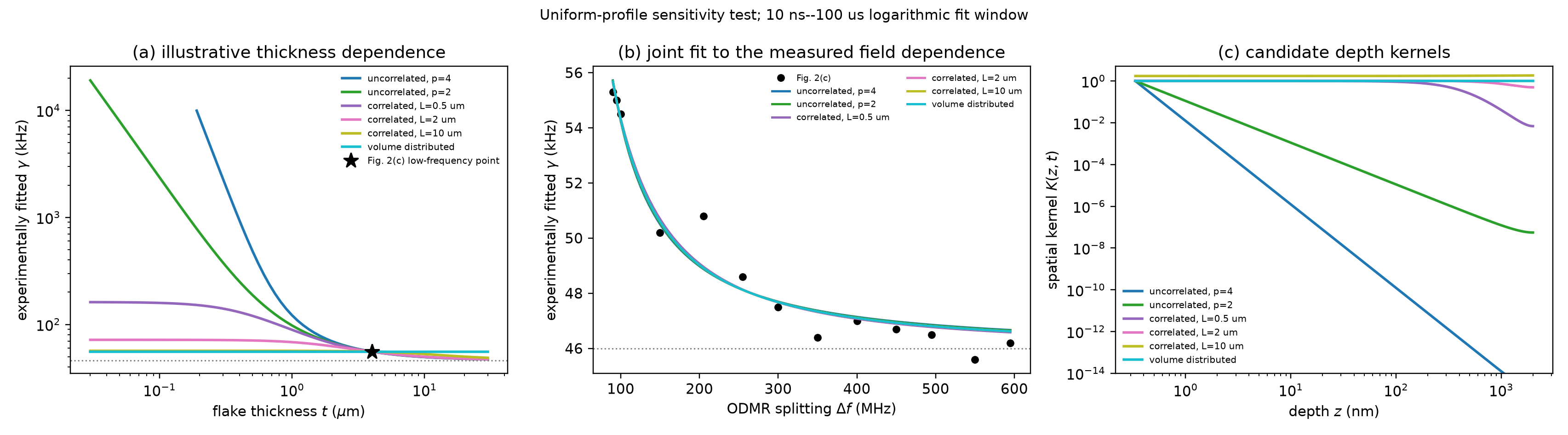}
  \caption{Sensitivity analysis for the candidate spatial models, using a uniform defect profile and the experimental fitting procedure of Eq.~\eqref{eq:dq-average}. (a) Illustrative thickness dependence at $\Delta f=90$~MHz. Portions for which at least half of the modeled ensemble has relaxed before the first delay are omitted because $\gamma$ is then not identifiable. (b) Joint fits to the field dependence in Fig.~\ref{fig:relaxation}(c). (c) Corresponding depth kernels. These curves are conditional predictions rather than measurements of the defect profile or the surface correlation length.}
  \label{fig:noise-model}
\end{figure*}

A thickness series can distinguish a volume source from a correlated surface only if it spans the relevant scale $L$. For $t\ll L$, the correlated kernel is nearly constant across the flake and is therefore almost degenerate with the volume model; measurements must extend to $t\gtrsim L$ to locate the crossover. The exponent $p$ also changes the crossover shape and cannot be absorbed into a prefactor. A measured irradiation profile and the recorded delay times should be used for each thickness before extracting a quantitative $L$.

\subsection{Expected dependence on defect density}

The present work uses one proton-irradiated sample at a fixed fluence, so it does not establish a density dependence. Increasing the $V_{\mathrm{B}}^{-}$ density may increase the photoluminescence signal, but it can also enhance defect--defect interactions and irradiation-induced strain or charge disorder. The resulting sensing optimum therefore requires a controlled density series.

The paired rates provide a useful diagnostic: a stronger change in $\gamma$ than in $\gamma_0$ would indicate an increased electric-field/strain channel, whereas a common change could include magnetic defect--defect interactions. This comparison would separate channel classes, but would not identify charge disorder independently of strain without an additional control such as surface treatment or gating \cite{zhou2025detection}.
\subsection{Limitations}

The quantitative values above are conditional on the uniform depth profile and on the representative kernels in Sec.~\ref{sec:noise-kernels}. A measured 250-keV irradiation profile could change the inferred local amplitudes and thickness curves. The present calculation also uses the logarithmic range shown in Fig.~\ref{fig:relaxation}(b); the recorded delay array and pointwise uncertainties should be supplied for a fully weighted reproduction of the experimental fit. Finally, $\gamma_{\infty}$ denotes only the component that is approximately field independent over the measured range. It may contain phonon, nuclear-spin, irradiation-induced, or spectrally flat surface contributions, which require temperature, density, or surface-control measurements to separate.

\end{document}